\documentclass[12pt]{iopart}
\usepackage{iopams,graphicx,color}
\usepackage{booktabs}

\usepackage{ulem}
\usepackage{textcomp}

\begin{document}

\title[Oscillations of a ball in a water jet]{Oscillations due to time-delayed driving of a ball in a water jet  - a challenging problem of the International Physicists' Tournament 2019 }

\author{S Michalke$^11$, A F\"osel$^2$, M Schmiedeberg$^3$}

\address{$^1$ Department Physik, Friedrich-Alexander-Universit\"at Erlangen-N\"urnberg, Staudtstr. 7, 91058 Erlangen, Germany}
\address{$^2$ Didaktik der Physik, Department Physik, Friedrich-Alexander-Universit\"at Erlangen-N\"urnberg, Staudtstr. 7, 91058 Erlangen, Germany}
\address{$^3$ Institut f\"ur Theoretische Physik 1, Friedrich-Alexander-Universit\"at Erlangen-N\"urnberg, Staudtstr. 7, 91058 Erlangen, Germany}
\ead{michael.schmiedeberg@fau.de}
\vspace{10pt}
\begin{indented}
\item[]  april 2020 
\end{indented}

\begin{abstract} The \textit{International Physicists' Tournament} (IPT) 2019 dealt with 17 challenging problems. In this article, we present experimental as well as theoretical approaches exemplarily to one of these tasks. A ball placed on a hard and flat surface can show oscillation movement when being hit by a jet of water from above. To explain the oscillations, a theoretical model is introduced and its predictions are compared to the experimental measurements. Furthermore, the structure of the IPT itself is characterized shortly in this article, as well as the idea of a new and innovative seminar which was set up at Erlangen University to prepare and assist students in taking part in International Physicists' Tournament. Furthermore, the educational relevance of physics tournaments like IPT for physics education at university is discussed in detail. 
\end{abstract}

\submitto{\EJP}

\maketitle

\section{Introduction}

The \textit{International Physicists' Tournament} (IPT) is a competition where physics students work on experiments, perform simulations or develop theoretical approaches in order to solve open problems. During the tournament the students have to present their solutions and discuss the approaches of other participants.
The characteristics of a (physics) tournament, especially in comparison to a non-tournament (physics) competition, are described in "International Physicists' Tournament - the team competition in physics for university students" by Vladimir Vanovskiy, member of the IPT executive committee \cite{vanovskiy}.

The German Physicists' Tournament (GPT) is a corresponding competition in Germany with problems from the upcoming IPT. The winning team of the GPT obtains the possibility to qualify for the IPT.

In order to support the students who participate in the competitions a new type of a seminar was developed at the Friedrich-Alexander-University (FAU) Erlangen-N\"urnberg, where students can both work on IPT problems and obtain credits for their studies. 

In this work, we will exemplarily present the experimental as well as theoretical approach to one of the problems of the IPT, namely problem number 5 from the 11th IPT that reads \cite{IPT}

\textit{
'When a ball lying on a hard and flat surface is hit by a jet of water that falls perpendicular to the surface, it may start to oscillate. Investigate how the oscillations depend on the relevant parameters.'
}

\begin{figure}[h]
\begin{center}
\includegraphics[width=.7\textwidth]{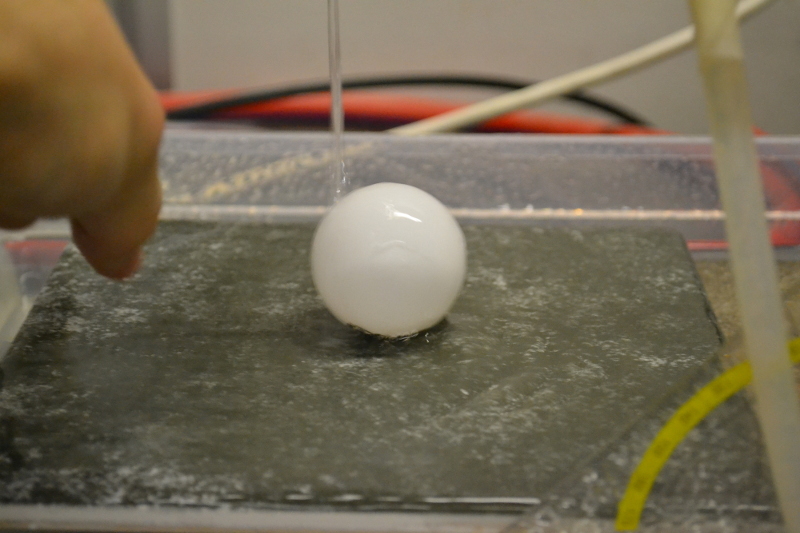}
\end{center}
\caption{A ping pong ball lying on a flat surface being hit by a water jet.\label{fig:oscillation_intro}}
\end{figure}

The situation is shown in Fig. \ref{fig:oscillation_intro}. The ball, in this case a ping-pong ball, lies on a stone slab and is hit by a jet of water. The water jet pulls the ball back into the middle and can induce a self-driven oscillation that will be studied in detail in this article.
Note that the oscillations are caused by an effective restoring force, which occurs for any displacement away from the resting position. However, a restoring force alone is not sufficient to keep oscillations alive in a system with friction. We will report on the time delay between the displacement and the restoring force that results in the driving of the oscillator. 

The article is organized as follows: After a short introduction to the IPT, the GPT and the related seminar in Sec. \ref{sec:ipt}, we discuss in Sec. \ref{sec:didactics} the educational relevance of those scientific tournaments for physics studies at university, which follow the specific ideas of the IPT. The theoretical and experimental results on the considered problem are presented in sections \ref{sec:description} to \ref{sec:experiment}. To be specific, we report our observations concerning the phenomena concerning the chosen problem in Sec. \ref{sec:description}. In Sec. \ref{sec:theory} we develop a theoretical description of the system. In Sec. \ref{sec:experiment} our theoretical predictions are compared to experiments. Finally, we conclude  and give an outlook on possible future developments of the competition in Sec. \ref{sec:conclusions}.

\section{International Physicists' Tournament, German Physicists' Tournament and preparatory seminar at Erlangen University}
\label{sec:ipt}
Since 2009 every year students can compete in the IPT. Within the months of june to august in the year before the international tournament, a set of 17 open problems is published that teams of up to six students can try to solve with experiments, simulations or theoretical calculations. The teams are welcome to puzzle over all of the 17 problems. According to the official rules of the international competition IPT \cite{IPTrules} a competing team can challenge some of the problems and each team has the possibility to reject some of the challenges. However, if a team rejects to many challenges it might be penalized. Therefore, it is recommended to prepare a total of about 13 problems for the IPT.
 
The problems usually deal with physical phenomena or questions that cannot be answered in a unique way but often allow for many possible ways to be handled. For example, the problems of IPT 2019 \cite{IPT} asked to build a radio with a potato (problem \textnumero{3}), to analyse in a non-invasive way whether the graphite rod of a pencil is broken (problem \textnumero{6}), to use the time-delay in a camera to measure the speed of light (problem \textnumero{16}) or to study the oscillations of a ball in a water jet  (problem \textnumero{5}). Exemplarily for all the challenging problems of IPT 2019, the latter phenomenon is discussed in this article (see sections \ref{sec:description} - \ref{sec:experiment}). At the actual tournament itself each round consists of so-called physics fights, where one student (termed ``reporter'') presents the results of the team concerning one of the problems. Afterwards, a student from another team called the ``opponent'' tries to criticize the reporting student. Finally, a student from a third team known as ``reviewer'' should comment on both the report as well as the criticism raised by the opposing student. In 2019, the IPT took place in april in Lausanne with 19 participating teams from 16 different countries. In order to qualify for that \textit{international} tournament a team has to be successful in an online qualification process. In some countries, there is a national competition instead and the winning team of such a national competition qualifies for the IPT. 
 
In Germany, the procedure mentioned last is practiced for finding the one team for representing Germany in the IPT, and the national competition is called \textit{German Physicists' Tournament} (GPT). It usually takes place in november or december in the year before the IPT. In december 2018 three teams competed in Frankfurt in order to qualify for the IPT 2019. The GPT 2019, which aimed to be the preselection for finding the team representing Germany at the IPT 2020, took place at the physics department of FAU Erlangen-N\"urnberg in december 2019 (for further information, see current website of German Physicists' Tournament \cite{fauGPT}). The GPT is based on the same rule set as the IPT with the sole modification that each team has to name the problems it has worked on (at least four per team) and that only these problems can be challenged.

While the IPT fosters the autonomy of the students and trains them to deal with complex challenges, the work on the problems usually also requires a lot of time which might be hard for students that already have to deal with the large workload of their physics studies. As a consequence, there are approaches to integrate the work on IPT problems as elective modules into the curriculum of physics education at the university. 
At the FAU Erlangen-N\"urnberg we created a new type of seminar, called 'Problems of the International Physicists' Tournament' at two credit hours and 5 European Credit Transfer System (ECTS) points, corresponding to a workload of about 140 hours. 

That seminar first took place in the winter term of 2018/2019 (mid of october 2018 until mid of february 2019), starting thus significantly after the publication of the IPT 2019-problems. Six students that later formed a team at the tournament participated in the seminar. To each student a supervisor was assigned, e.g. a PhD student interested in the tournament.
The task of the supervisors was to help the students with the theoretical background of the problems, e.g. by finding literature, explaining typical approaches in theoretical modelling, as well as by helping with code development if simulations were employed. Furthermore, the supervisors enabled contacts to other researchers at the university that are closer to the field of the problem. Finally, contacts to the practical training courses were supported. The supervisors worked for the seminar as part of their mandatory teaching load. In addition, they benefited from obtaining insides into how physical questions can be attacked in various fields of physics, including questions that are outside of their own field of research.

The seminar consisted of two elements: First, the participating students presented their approaches as well as the physical background concerning one IPT problem of their choice in talks of 45 minutes.
Approaches to a task included various ideas how a problem can be solved. Not all of these ideas had to be worked out in detail. Usually one or two those ideas were followed in more detail, e.g. by presenting results of preliminary experiments or simulations. However, not a worked-through, single solution in the style of a presentation in a physics fight had to be presented, but the students had to show that a problem can be approached from various directions.
The talks were followed by an extensive discussion with all members of the audience that not only consisted of the other students, but also of other members of the university that might be related to the respective field of the problem. The purpose of these talks is that the students not only learn about possible approaches to solve the considered problems, but that they also discuss related phenomena and their explanations if known in literature. Due to these talks the seminar could be graded and counted with 5 credits. The grades were given by the professor after a brief discussion with the respective supervisor. Therefore, the effort of the students for puzzling over the IPT-problems is well appreciated.

Second, training fights according to the rules of the IPT took place such that the students could discuss their approaches to the problems in a similar way as at the IPT. The training fights were not mandatory for the seminar participants, but the students enjoyed to engage and to prepare for the competition.

Note that in each meeting of the seminar, there first was a talk and then a fight with other students. Therefore, some students first had a talk in the seminar and a few weeks later their physics fight presentation. Other students first presented one possible solution in a physics fight and later had to present the problem in a talk from a broader perspective. In both cases the students were able to include the feedback on their first presentation or talk when they took the stage for the second time. The tight schedule was chosen because all talks and all test fights should be scheduled between the start of the semester in october and the date of the GPT that took place in december. Therefore, at GPT, the six participants of the seminar had well prepared six problems which is sufficient for the GPT. However, for the IPT the students had to work on additional problems which turned out to be difficult without the full support of the seminar. Therefore, in future seminars each student will give a talk and a short presentation for a physics fight not about the same, but about different problems. To prevent a too tight schedule, future seminars are planned to already start in the summer term. 

The seminar was repeated in the winter term 2019/2020 and will take place again in 2020/2021, but this time we will already start in the second half of the summer term (april 2020 to july 2020) and then be continued in the first half of the winter term 2020/2021: Students should already be familiar with discussing open problems in june or july, so that right before the student summer holidays in august the team(s) for GPT could be fixed. Of course, there might be the problem, that at this time the tasks of IPT 2021 won't be published, but for just getting familiar with the structure of IPT-problems, one could work on IPT-problems from earlier years. Based on our experiences with the previous seminars, the time spent with training fights probably will be reduced, i.e., training fights will mainly be performed right before the GPT. Instead we want to enable more free discussions about possible ways how to work on the IPT problems. This might include short presentations on ideas for experiments, simulations or calculations. That way, more problems can be discussed and the exchange between the students is intensified.

\section{Educational relevance of physics competitions for physics education}
\label{sec:didactics}

Though there don't exist any studies at all investigating the effectiveness of using physics competitions for physics education \textit{at university}, we definitively think that there are quite a few arguments for including physics competitions in physics education at university. This is especially true for those so-called physics tournaments which are dealing with creative, innovative and challenging open tasks and which are asking for qualifications and competencies far beyond just only specialised knowledge in physics. Subsequently, we will present and illustrate some of the arguments.

Including physics competitions in science education allows for taking into account the methodical and didactic ideas of \textit{active learning} (see e.g. \cite{shepard}, \cite{diamond}). The following text passage may illustrate the main ideas: In 1988, the American theoretical physicist \textit{Richard Feynman} gave an account of his interactions with his father in the essay "How to become a Scientist?" published in his book "What do You care what other people think? Further adventures of a curios character."\cite{feynman} Feynman's father taught him \textit{to carefully observe phenomena}, and he triggered him to ask questions arising from an observation (e.g. "Why do you think birds pick at their feathers?"). Feynman's father, he never gave the answer immediately, but rather he would get Feynman \textit{to think of} an explanation, to devise an experiment or to provide an explanation - \textit{to understand the meaning of the phenomena}. So the point was not in giving the answer, but rather in getting Feynman to learn "the difference between knowing the name of something and knowing something" (\cite{feynman}, p. 14) by being active in learning. Thus active learning in the context of physics tournaments like IPT with all the open problems mostly arising from observations from daily live means that students make the observations by themselves, e.g. they carefully look at the oscillations of the driven ball. They have to design experiments in a self-contained way as well as they have to provide appropriate theories, and they have to find possible solutions. In the contest itself, they present and defend their own ideas and solutions. 

(Science) Competitions \textit{provide students with an opportunity to evaluate their performance with others'}. Student competitions can thus play a crucial role in identifying talented students. This is quite a good possibility for students at university to get to know their own subject-specific competencies (being talented in doing experiments or in making theoretical approaches) as well as to find out something about competencies concerning soft-skills (e.g. when getting an award for "best presenter" or "best opponent"). And this is not only a chance for students, but for researchers at university, too: Via competitions they get to know talented and/ or highly motivated students which they can encourage to take part in research studies at their department. 

The empiricist \textit{Campbell} stated, that participation in science competitions helps students \textit{become aware of their potential} and \textit{contributes to their self-confidence} (see \cite{campbell 1996}, \cite{campbell 2002} and \cite{campbell 2002a}). This is a statement about students in schools; it might be true as well for students in university, though empirical studies are still missing and without studies there is no evidence that one can infer from one population (students in schools) to another (students at university).

Various interest studies have shown that physics competitions \textit{increase the interest and the motivation} of students in schools (see \cite{motivation1}, page 7-15, \cite{motivation2} and \cite{motivation3}), at least of those students who have been successful.

Integrating IPT in physics education pushes the handling with scientific methods, and 
scientific methods like "observing", "measuring", "testing hypotheses", "modelling", "experimenting"
and "interpreting" help with learning concepts and principles of science (\textit{learning OF science}). Getting to know scientific methods helps understanding the nature of science (\textit{learning ABOUT science}) and scientific methods are a way towards "goals" not specific for just one subject, e.g.
learning how to solve problems or acquiring critical faculties (\textit{DOING science}). While
solving the open problems, students apply different scientific methods and therefore learn OF science,
they also learn ABOUT science and - most important - they DO science. A wide range of
literature about the subjects "scientific methods and science education" and "nature of science"
have been published. \textit{Wynne Harlen}, for example, discusses scientific methods in the context of
teaching, learning and assessing science (\cite{Harlen 2006}). \textit{McComas} (\cite{Mc Comas 1998}),
\textit{H\"ottecke} (\cite{Hoettecke 2001}) and \textit{Neumann} (\cite{Neumann 2011}) introduce in the nature of science.

Presenting, discussing and defending the problems in the contest itself finally fosters (self-)criticism as well as competencies concerning communication and decision-making.

\section{Qualitative description of the phenomena}
\label{sec:description}

\begin{figure}[h]
\begin{center}
(a)\includegraphics[width=0.47\textwidth]{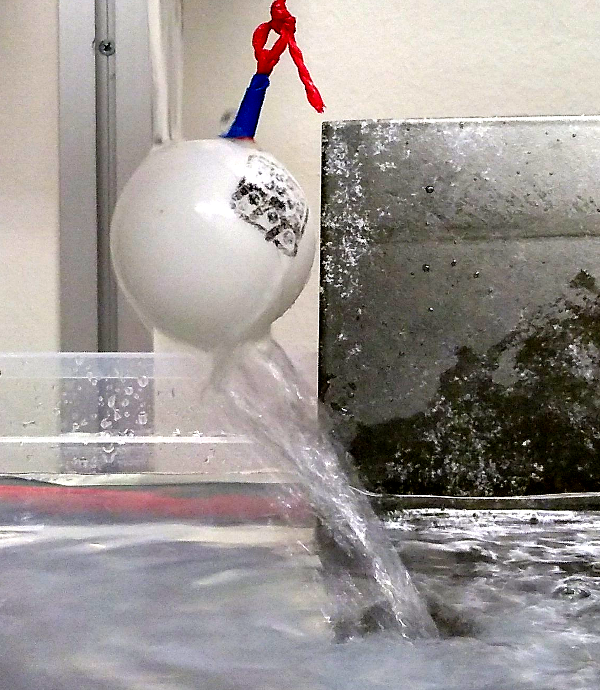}
\ \ \ \ \ 
(b)\includegraphics[width=0.38\textwidth]{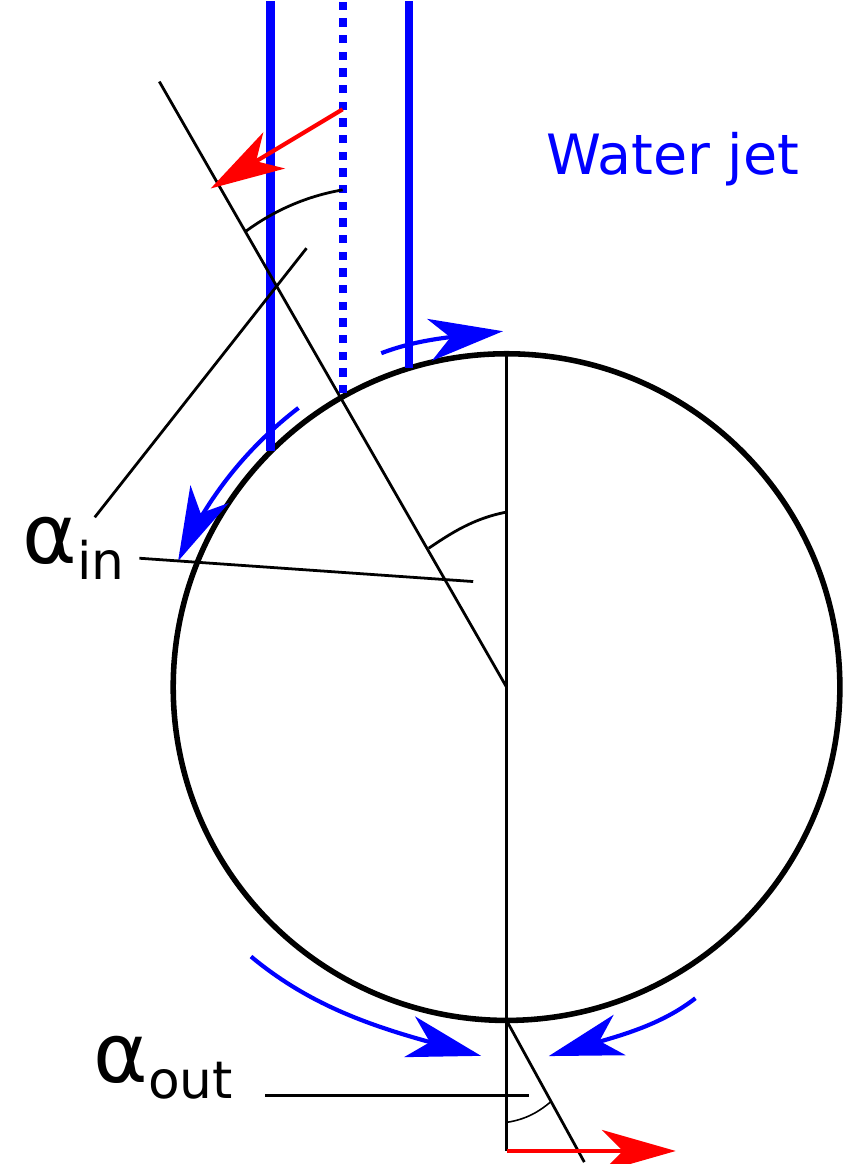}
\end{center}
\caption{(a) The water of a jet flows along the surface of a hanging ball. If the water hits the ball off center, a new water jet forms at the bottom of the ball and leaves the ball in a direction opposite to the side where ball was hit by the initial jet. (b) Geometry of the incoming and outgoing water jet. The angle of the incoming jet is termed $\alpha_{\mathrm{in}}$ and of the outgoing one $\alpha_{\mathrm{out}}$. Note that in the stationary case we observe $\alpha_{\mathrm{in}}=\alpha_{\mathrm{out}}$. However, in case of a moving ball, the $\alpha_{\mathrm{out}}$ is time-delayed with respect to $\alpha_{\mathrm{in}}$, i.e., $\alpha_{\mathrm{out}}(t)=\alpha_{\mathrm{in}}(t-\Delta t)$ with a delay time $\Delta t$. \label{fig:jet_bending}}
\end{figure}

We observe three qualitative different behaviors of the ball that is hit by the jet of water from above: First, damped oscillations can be seen in case of a weak water jet. The resting position of the oscillations is the position where the ball is hit from the top in a symmetric way, i.e., where the jet impacts above the center point of the ball. If the ball is displaced from the resting position there is a restoring force acting towards the resting position. Due to the damping the oscillations are damped and for weak water jets there is no driving that is sufficient to uphold the oscillations. Second, in case of an increased water flow, the oscillations persist due to a driving mechanism that we will discuss in this article. Third, if the water jet becomes too strong it is deflected from the ball and might splash in different directions. Thus the motion of the ball becomes chaotic and as a consequence hard to predict.

Since the task of the competition concerns the oscillations, we especially studied the first and second case, where either damped or persisting oscillations are observed. The water from the jet flows along the surface of the ball until it hits the surface where the ball is placed on.

In order to study how the water flows along the surface of the ball, we consider a hanging ball that is targeted by a vertical water jet in a slightly asymmetric way, i.e., the water jet does not hit the ball directly on top but a short distance besides (see Fig. \ref{fig:jet_bending}(a) for a photograph and Fig. \ref{fig:jet_bending}(b) for a sketch). The water then flows along all sides of the ball and forms a new jet at the bottom of the ball. Interesting, the outgoing water jet does not leave the ball in vertical direction but is clearly bent into the direction that is opposite to the side where the incoming jet has hit the ball.

In this article we will explain that the momentum carried by the outgoing water jet is essential for understanding the driving mechanism. Since the water jet is deflected at the ball, there is a momentum transfer onto the ball which leads to the restoring force towards the resting position of the oscillations. We argue that the driving is due to a time delay of the restoring force as we will explain in Sec. \ref{sec:theory}. Note that we assume that a similar momentum is transferred into the water layer if the ball is placed onto a surface.

\section{Theory of the Oscillator}
\label{sec:theory}

\subsection{Equation of motion with velocity-dependent restoring force}

The motion of a ball that is displaced in an arbitrary direction in parallel to the surface is described by the equation
\begin{equation}
m \ddot{\vec{r}}(t) = \vec{F}\left(\vec{r}(t), \dot{\vec{r}}(t)\right)  - \gamma \dot{\vec{r}}(t),
\end{equation}
where $\vec{r}(t)$ is the displacement of the center of the ball from its resting position, $m$ is the mass of the ball or, if rotation is involved, a term that denotes the inertia in general, $\vec{F}\left(\vec{r}(t), \dot{\vec{r}}(t)\right)$ is the restoring force that in the harmonic approximation would be $ \vec{F}\left(\vec{r}(t), \dot{\vec{r}}(t)\right)=-k \vec{r}(t)$, and $\gamma \dot{\vec{r}}(t)$ with a friction coefficient $\gamma$ is the damping force.

Note that we neglect contributions due to the rotation of the ball because we do not observe any rapid rotations. Slow rotations as for a ball that is not just displaced out of the resting position but that is rolled out by a small angle do not change the qualitative form of the equation of motion in case of small displacements from the resting position. 

The restoring force mainly originates from the deflection of the water jet that in case of a ball that is fixed with a certain displacement $\vec{r}$ but with no velocity (i.e., $\dot{\vec{r}}=\vec{0}$) leads to a force $\vec{F}_0$ that depends on the absolute value of the displacement, i.e.,
\begin{equation}
\vec{F}_0\left(\vec{r}\right)=\vec{F}\left(\vec{r},\dot{\vec{r}}=\vec{0}\right)=-f\left(\left|\vec{r}\right|\right)\frac{\vec{r}}{\left|\vec{r}\right|}.
\end{equation}
Again, in a harmonic approximation $f\left(\left|\vec{r}\right|\right)=k\left|\vec{r}\right|$, which we will later assume for simplicity in our more detailed analysis.

\begin{figure}[h]
\begin{center}
\includegraphics[width=0.3\textwidth]{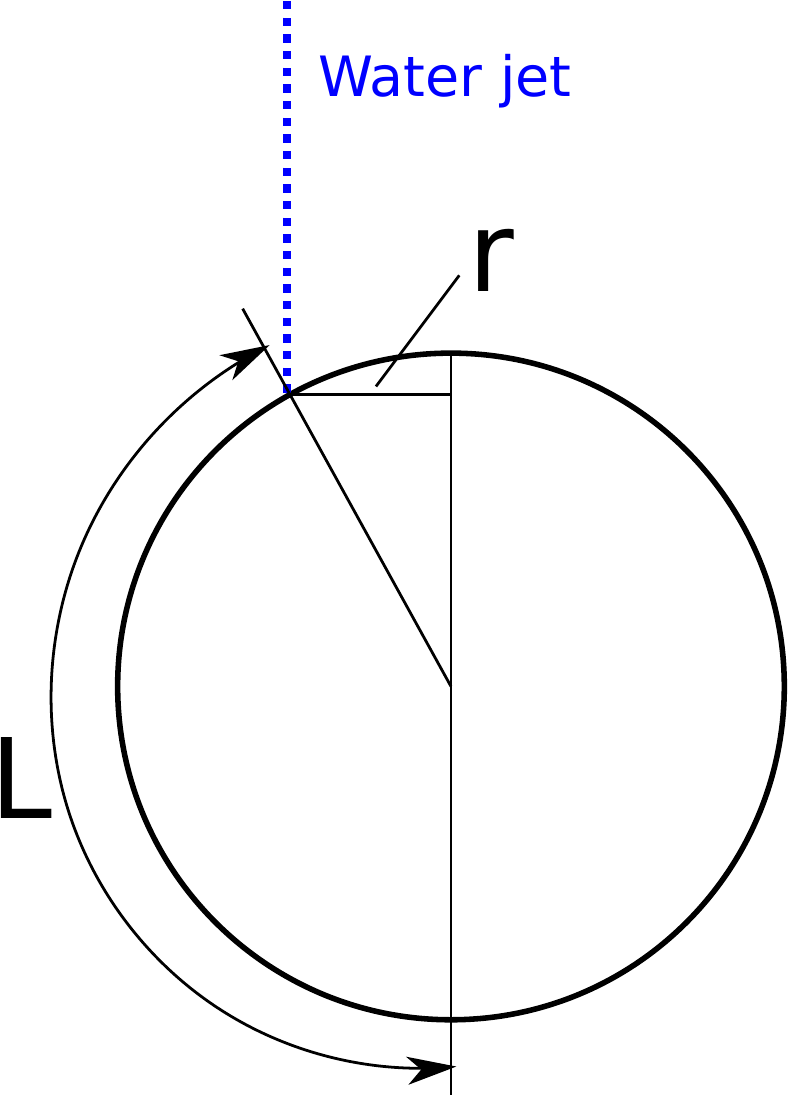}
\end{center}
\caption{The outgoing jet that leaves the ball at the bottom is delayed with respect to the incoming jet.  The delay can be estimated by using the delay length $L\left(\vec{r}(t)\right)$ and the typical velocity $v_{\textrm{w}}$ of the water flowing around the ball.\label{fig:delay_distance_descr}}
\end{figure}

During the oscillation, the ball is obviously never at rest for any constant displacement $\vec{r}$. Therefore, we have to think about how the force $\vec{F}\left(\vec{r}(t), \dot{\vec{r}}(t)\right)$ depends on the velocity $\dot{\vec{r}}(t)$. In order to motivate such a velocity-dependent force, we take another look on Fig. \ref{fig:jet_bending}(b). The restoring force is due to the deflection of the water jet. To be specific, the momentum that is exerted on the ball in horizontal direction is opposite to the momentum that is carried away by the outgoing jet. Note however, that the outgoing water jet is delayed with respect to the incoming jet. The delay can be roughly estimated by a delay length $L\left(\vec{r}(t)\right)$ as sketched in Fig. \ref{fig:delay_distance_descr} and the typical velocity $v_{\textrm{w}}$ of the water flowing around the ball, i.e., the time delay is
\begin{equation}
\Delta t = \frac{L\left(\vec{r}(t)\right)}{v_{\textrm{w}}}.
\end{equation}
As a consequence, the time-delayed force approximately is
\begin{equation}
\vec{F}\left(\vec{r}(t),\dot{\vec{r}}(t)\right)\approx\vec{F}_0\left(\vec{r}(t)-\Delta t\dot{\vec{r}}(t)\right)=\vec{F}_0\left(\vec{r}(t)-\frac{L\left(\vec{r}(t)\right)}{v_{\textrm{w}}}\dot{\vec{r}}(t)\right).
\end{equation}
For this approximation, we assumed that the time delay is small such that the velocity $\dot{\vec{r}}(t)$ can be considered approximately constant during the time delay. As we will show in the following, this time delay is the origin of the driving of the oscillations.

In the harmonic approximation the equation of motion with a time-delayed harmonic restoring force is
\begin{eqnarray}
  \nonumber
  m \ddot{\vec{r}}(t) &= - k\left(\vec{r}(t)-\frac{L}{v_{\textrm{w}}}\dot{\vec{r}}(t)\right)  - \gamma \dot{\vec{r}}(t)=- k\vec{r}(t)+\left(k\frac{L\left(\vec{r}(t)\right)}{v_{\textrm{w}}}- \gamma\right)\dot{\vec{r}}(t)\\
  &=- k\vec{r}(t)-\gamma_{\textrm{eff}}\left(\vec{r}(t)\right)\dot{\vec{r}}(t).
\end{eqnarray}
In the last step we introduced an effective friction constant $\gamma_{\textrm{eff}}\left(\vec{r}(t)\right)=\gamma-k\frac{L\left(\vec{r}(t)\right)}{v_{\textrm{w}}}$. Therefore, the equation of motion corresponds to an oscillator with a harmonic restoring force $- k\left(\vec{r}(t)\right)$ and a damping contribution $-\gamma_{\textrm{eff}}\left(\vec{r}(t)\right)\dot{\vec{r}}(t)$. The effective friction constant $\gamma_{\textrm{eff}}\left(\vec{r}(t)\right)$ can depend on the position, such that the equation becomes nonlinear. Furthermore, the $\gamma_{\textrm{eff}}\left(\vec{r}(t)\right)$ might be positive indicating that there is an overall driving instead of pure dissipation.

Finally, the delay length $L\left(\vec{r}(t)\right)$ is given by (see Fig. \ref{fig:delay_distance_descr}):
  \begin{equation}
L\left(\vec{r}(t)\right)=R\left[\pi -\arcsin\left(\left|\vec{r}(t)\right|/R\right)\right]\approx \pi R-\left|\vec{r}(t)\right|,
\end{equation}
where $R$ is the radius of the ball. Therefore, the equation of motion is
\begin{equation}
\label{eq:complete_diff}
m \ddot{\vec{r}}(t) = - k\vec{r}(t)-\gamma_{\textrm{eff}}\left(\vec{r}(t)\right)\dot{\vec{r}}(t)
\end{equation}
with
\begin{equation}
\label{eq:gamma-exakt}
  \gamma_{\textrm{eff}}\left(\vec{r}(t)\right)=\gamma-k\frac{\pi R-\arcsin\left(\left|\vec{r}(t)\right|/R\right)R}{v_{\textrm{w}}}
\end{equation}
or as an approximation
\begin{equation}
\label{eq:gamma-approx}
 \gamma_{\textrm{eff}}\left(\vec{r}(t)\right) \approx\gamma-k\frac{\pi R-\left|\vec{r}(t)\right|}{v_{\textrm{w}}}.
\end{equation}

\subsection{Trajectories of the oscillations in phase space}

\begin{figure}[h]
\begin{center}
        \includegraphics[width=0.3\textwidth]{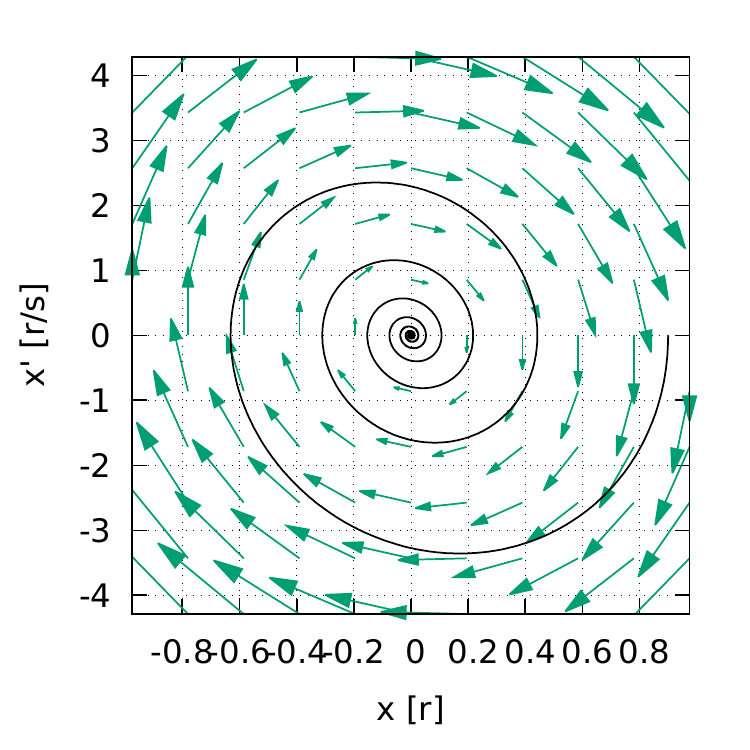}
        \includegraphics[width=0.3\textwidth]{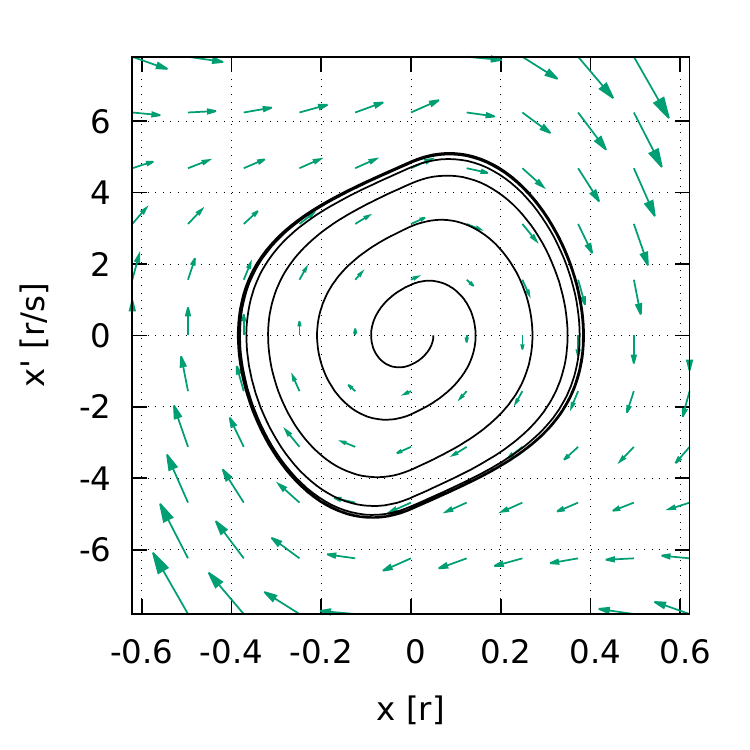}
        \includegraphics[width=0.3\textwidth]{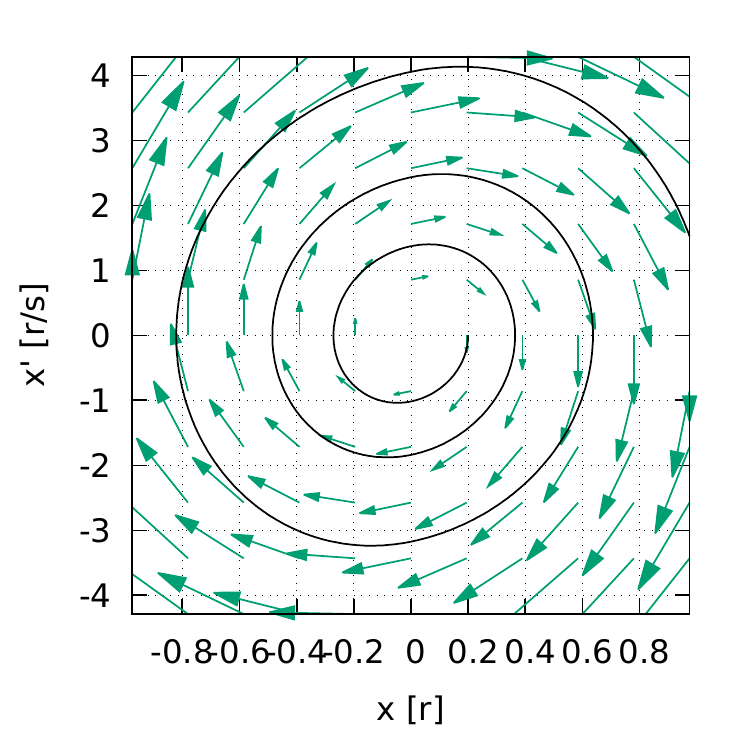}\\
        \hspace{0.027\textwidth} (a) \hspace{0.26\textwidth} (b) \hspace{0.26\textwidth} (c)
\end{center}

\caption{(a) Trajectory of a damped oscillation in phase space. The trajectory starts at $\vec{r}(0)=(x_0,0)$ with $x_0=0.9 R$ and zero velocity. (b) Trajectoriy converging against an limiting cycle. The trajectories are started with $x_0=0.05 R$. (c)  Trajectory of an oscillator where the driving is stronger than the damping. The trajectory starts with $x_0=0.2 R$ and zero velocity. The parameters of the simulation are $\frac{k m}{\gamma^2} = 19.34$ in all cases and (a) $R\gamma/(v_{\textrm{w}}m) = 10.0\cdot 10^{-4}$, (b) $R\gamma/(v_{\textrm{w}}m) = 8.4\cdot 10^{-4}$ and (c) $R\gamma/(v_{\textrm{w}}m) = 7.0\cdot 10^{-4}$.}

 \label{fig:phase_space_1}
 \end{figure}

Equations (\ref{eq:complete_diff}) and (\ref{eq:gamma-exakt}) can be used to simulate the trajectory numerically. This has been done using Runge Kutta Method of 4th order. In Fig. \ref{fig:phase_space_1} the trajectory of a damped and unbound oscillation is shown in black color. The derivatives are indicated by green arrows.

\subsection{Estimation of the amplitude in case of self-sustaining oscillations}

In the following, we estimate the amplitude of oscillations in case of a steady state where the work added by the driving is equal to the dissipated work.

In order to determine the work we integrate over half a period
\begin{equation}
  \label{eq:half_period_int}
  W = \int_0^{\pi/\omega} m\ddot{\vec{r}}(t)\dot{\vec{r}}(t) dt
\end{equation}
using an harmonic oscillation as a rough estimate:
\begin{eqnarray}
\vec{r}(t) &= \vec{A} \cos(\omega t),
\label{eq:harm_assumption_1}\\
\dot{\vec{r}}(t) &=- \vec{A}\omega \sin(\omega t),
\label{eq:harm_assumption_2}
\end{eqnarray}
where $\vec{A}$ is the steady state amplitude and $\omega$ the circular frequency.
Using Eqs. (\ref{eq:complete_diff}) and (\ref{eq:gamma-approx}) one finds
\begin{eqnarray}
  \nonumber
  W &= \int_0^{\pi/\omega}\left[-k\vec{r}(t)-\left(\gamma-k\frac{\pi R-\left|\vec{r}(t)\right|}{v_{\textrm{w}}}\right)\dot{\vec{r}}(t)\right] \dot{\vec{r}}(t) dt\\
  \label{eq:w}
    &=A^2m\omega\left[\frac{\pi}{2}\left(\frac{k\pi R}{v_{\textrm{w}}}-\gamma\right)+\frac{2}{3}A\frac{k}{v_{\textrm{w}}}\right].
\end{eqnarray}

A self-sustained oscillation is achieved if the energy does not change, i.e., if the work integrated over half a period vanishes. For $W=0$ in Eq. (\ref{eq:w}) one obtains an estimate for the amplitude of a self-sustained oscillation
\begin{equation}
  \label{eq:ampl}
  A=\frac{3\pi}{4}\left(\pi R-\frac{\gamma v_{\textrm{w}}}{k}\right).
\end{equation}

Note that the estimate given in this section is only valid for sufficiently small amplitudes due to the approximation used in Eqs. (\ref{eq:gamma-approx}), (\ref{eq:harm_assumption_1}), and (\ref{eq:harm_assumption_2}).

\section{Experimental verification}
\label{sec:experiment}

In order to verify the calculations experiments were performed. First we investigate the frequency dependence on the properties of the water jet.

\subsection{Water jet characterization}

The water jet is characterized by the height $h_w$, the volume flow per time unit $\frac{\Delta V}{\Delta t}$ and the nozzle diameter $d_N$. Using the mass flow and the nozzle diameter the exit velocity of the water jet $v_{exit} = \frac{\Delta V}{\Delta t} \frac{1}{d_N^2 \pi}$ can be computed. The height between exit nozzle and the ball can then be used to derive the velocity from the water jet when hitting the ball $v_w$:
\begin{equation}
v_w  = \sqrt{v_{exit}^2 + 2\,g\,h_w} = \sqrt{\left( \frac{\Delta V}{\Delta t} \frac{1}{d_N^2 \pi} \right)^2 + 2\,g\,h_w} \, .
\end{equation}
$g$ denotes the gravitational constant. Multiplying the final water velocity $v_w$ with the volume flow and the density of the liquid $\rho$ results in the force of the water jet $F_w$ or equivalently the momentum per time:
\begin{equation}
F_w = v_w  \frac{\Delta V}{\Delta t} \rho \, .
\end{equation}
Using the assumption made in Fig. \ref{fig:jet_bending} we expect the jet to bend in a $90^\circ$ angle if the ball is at the position $|r|=R$. This results in the full momentum transfer. The constant of proportionality $k$ in the harmonic force approximation of the restoring force therefore is
\begin{equation}
k = \frac{F_w}{R} \, .
\label{eq:k_definition}
\end{equation}

\subsection{Experimental setup}

\begin{figure}[h]
\begin{center}
\includegraphics[width=.7\textwidth]{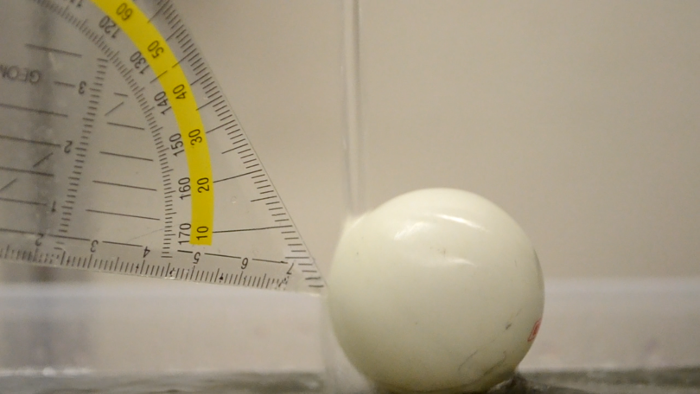}
\end{center}
\caption{The view of the camera.\label{fig:camera_view}}
\end{figure}

The setup consists out of a flat slab with a ping pong ball on top being hit by a water jet. The movement is tracked using a camera looking from the side onto the jet. The position is calibrated using the size of the ball which is 40$\,$mm. The view of the camera is shown in Fig. \ref{fig:camera_view}.

In order to excite oscillations of the ball a ruler is pushed onto the side and then removed quickly enough to allow the ball to oscillate freely. The motion of the ball then either is damped until it comes to rest or in case of sufficient driving it continues oscillating with a specific amplitude.

\subsection{Data fitting and extraction of oscillation parameters}

\begin{figure}[h]
\begin{center}
\includegraphics[width=.6\textwidth]{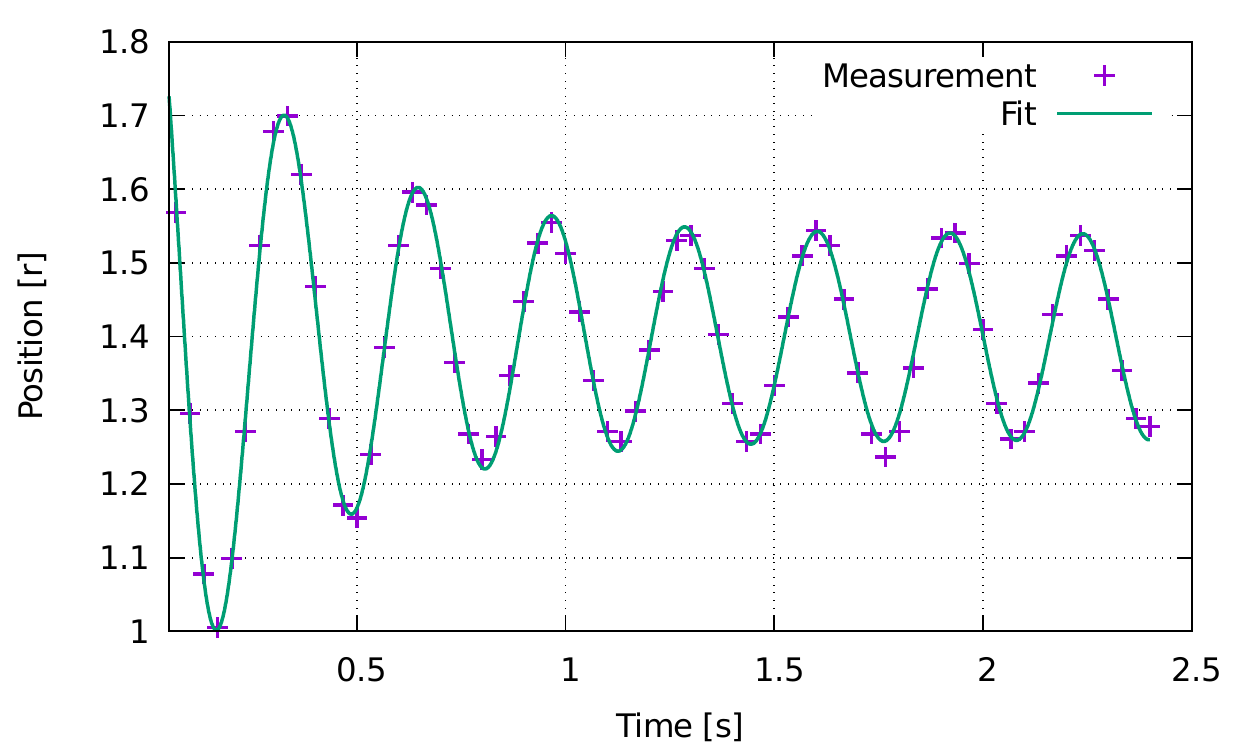}
\end{center}
\caption{Tracked positions during an oscillation (violett) and a fit (green) according to Eqs. (\ref{eq:fit1}), (\ref{eq:fit2}), and (\ref{eq:fit3}).\label{fig:eye_balled_fit}}
\end{figure}

The video is analyzed using ``OpenCV v2'' \cite{opencv}. The following function was fitted to the measurement data:
\begin{equation}
  \label{eq:fit1}
y(x) = A(x) \sin( 2 \pi f(x) \cdot (x+x_{off}) ) + y_{off},
\end{equation}
Where the amplitude and frequency approach a constant value according to an exponential ansatz
\begin{equation}
  \label{eq:fit2}
A(t) = (\mathrm{A}_0-\mathrm{A}_{\infty}) \cdot e^{-\gamma_f t} + \mathrm{A}_{\infty} 
\end{equation}
and 
\begin{equation}
  \label{eq:fit3}
f(t) = (f_0-f_{\infty}) \cdot e^{-\gamma_f t} + f_{\infty} \, .
\end{equation}
 The result of fits is shown in Fig. \ref{fig:eye_balled_fit}. Note that in principle the plane of oscillation might slightly change which would require an additional correction for long runs.

The asymptotic amplitude $\mathrm{A_{\infty}}$ that is approached for $t\rightarrow\infty$ corresponds to the steady state amplitude calculated in Eq. (\ref{eq:ampl}) and depends on the product of $\gamma$ and $v_{\mathrm{W}}$. Therefore, for a measured $\mathrm{A_{\infty}}$ the product of $\gamma$ and $v_{\mathrm{W}}$ can be determined:
\begin{equation}
\gamma v_\mathrm{W} = k \left( \pi R - \frac{4}{3} \frac{A_{\infty}}{\pi} \right).
\end{equation}

In order to determine $\gamma$ and $v_\mathrm{W}$ individually, we compare our experimental results to the results of the numerical simulation that was used to generate the plots in Fig. \ref{fig:phase_space_1}.

\begin{figure}[h]
\begin{center}
\begin{center}
        \includegraphics[width=0.49\textwidth]{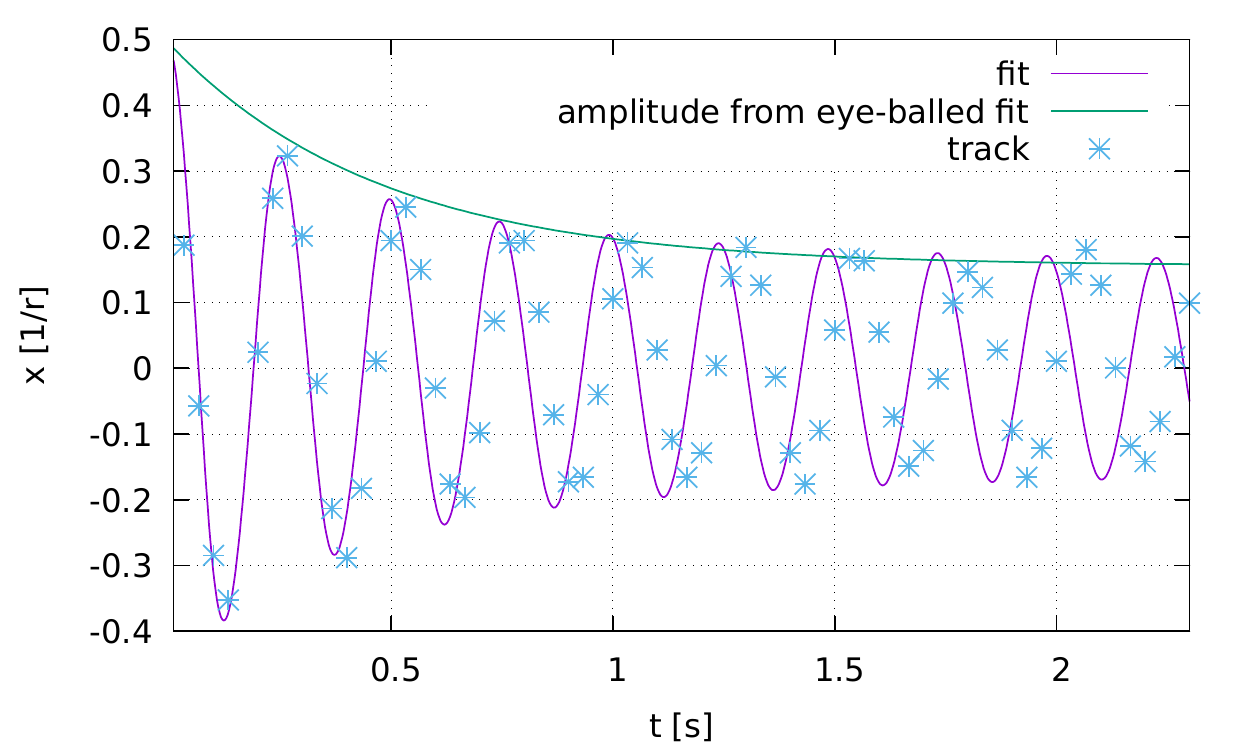}
        \includegraphics[width=0.49\textwidth]{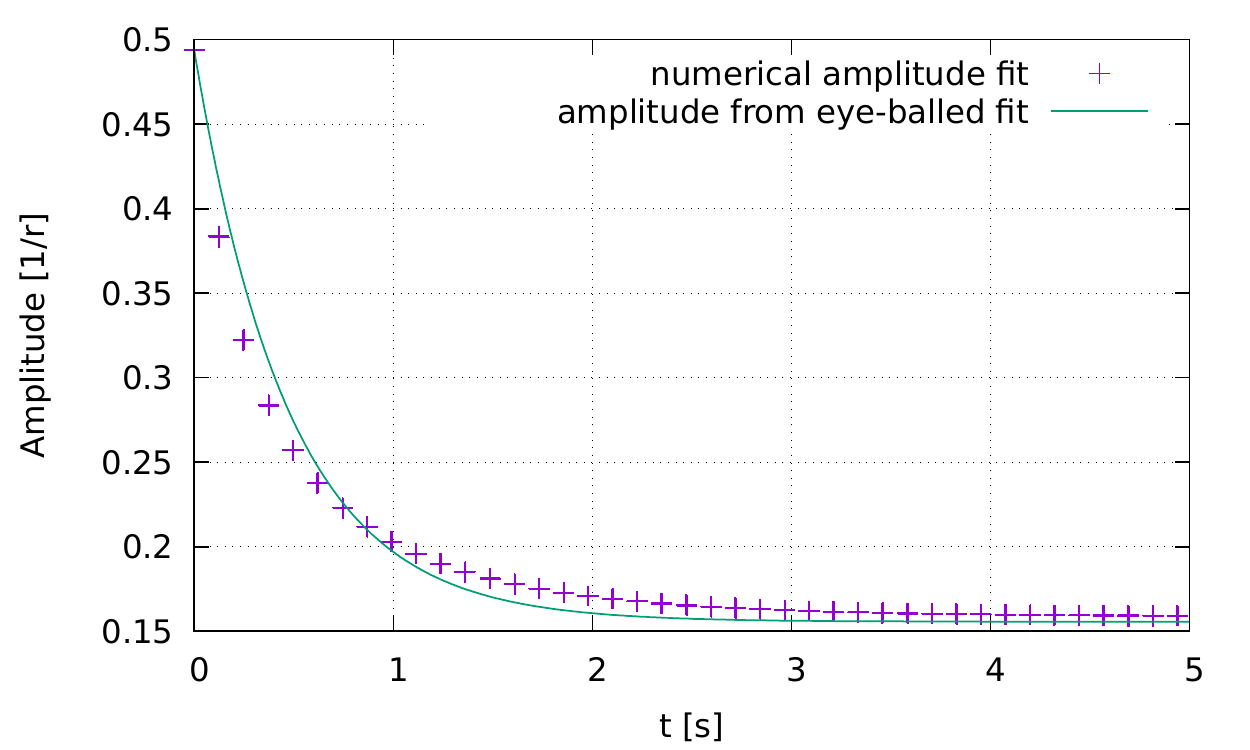}\\
        \hspace{0.047\textwidth} (a) \hspace{0.45\textwidth} (b)
\end{center}
\end{center}
\caption{(a) Comparison of the results from numerical simulations (stars) and the phenomenological fit curve to experimental data (magenta line). The parameters of the numerical simulation are varied until the deviations between the numerical extrema and the envelop function are minimal. (b) Example of the amplitudes obtained from a fit to simulations (crosses) and the amplitudes according to the fit to the experimental results (line). The results are similar, but from (a) and (b) it is clearly visible that the simulation does not implement slipping and therefore does not fit perfectly. Most important, the frequency change in the experiment in (a) is not reproduced by the numerical simulations without slip.
\label{fig:fit_explanation}}
\end{figure}

Starting with the phenomenological fit to the results of the experiment (see Fig. \ref{fig:eye_balled_fit}) we extract the decay of the envelope function. Then, we performed numerical simulations with different parameters, determined the maximum and minimum of the simulation results and compare their position to the envelope function of the fit to the experiment. To be specific, we vary $v_w$ and $\gamma$ in the numerical simulations until the sum of the differences between the simulated extrema and the experimental envelope function are minimal. In Fig. \ref{fig:fit_explanation}(a) the comparison of the optimized simulations and the experimental results are shown.
\color{black}

In table \ref{tab:fit_result} the parameters of simulations are shown that best fit the experiments.
\begin{figure}[h]
\begin{center}
\begin{tabular}{c||c|c|c|c|c}
$m_e$ [g] & $F_\mathrm{w}$ [mN] & $\gamma$ [1/$s$] & $v_\mathrm{w}$ [$r/s$] & $\gamma \cdot m_e$ $\left[\frac{\mathrm{mN}}{(r/t)}\right]$\\
\midrule
2.679$\pm$0.005 & 53.0$\pm$1.3 & 118$\pm$10 & 10.8$\pm$1.0& 0.53$\pm$0.05 \\
1.864$\pm$0.003 & 53.0$\pm$1.3 & 168$\pm$20 & 10.6$\pm$1.3 & 0.52$\pm$0.07 \\
2.679$\pm$0.005 & 31.3$\pm$0.5 &  82$\pm$13 &  6.6$\pm$1.0 & 0.37$\pm$0.06 \\
2.679$\pm$0.005 & 73.1$\pm$2.1 & 106$\pm$18 & 21.0$\pm$3.4 & 0.47 $\pm$0.08
\end{tabular}
\end{center}
\caption{The resulting values of simulations that best fit the results from experiments. $m_e$ denotes the mass of the ball, $F_\mathrm{w}$ the water force, $\gamma$, $v_\mathrm{w}$ the water velocity on the surface of the ball, and $\gamma \cdot m_e$ the water drag force.\label{tab:fit_result}}
\end{figure}

Using those values it is possible to do some plausibility checks of the theory. The first expected behavior is that the drag force $\gamma \cdot m_e$ and the water velocity $v_w$ should only depend on the geometry of the setup, the water velocity and the mass flow rate.  This is confirmed when looking at the first two rows in table \ref{tab:fit_result}. The water force was varied using higher or lower flow with the same nozzle and therefore a faster or slower water jet at the exit of the nozzle. This can also be confirmed with looking at the water velocity on the surface of the ball $v_w$. Interestingly, the drag force decreases for larger water force in the last row. This can be explained by the ball starting to lift up and slip which results in having less friction.

\subsection{Frequency observation}

\begin{figure}[h]
\begin{center}
\includegraphics[width=.7\textwidth]{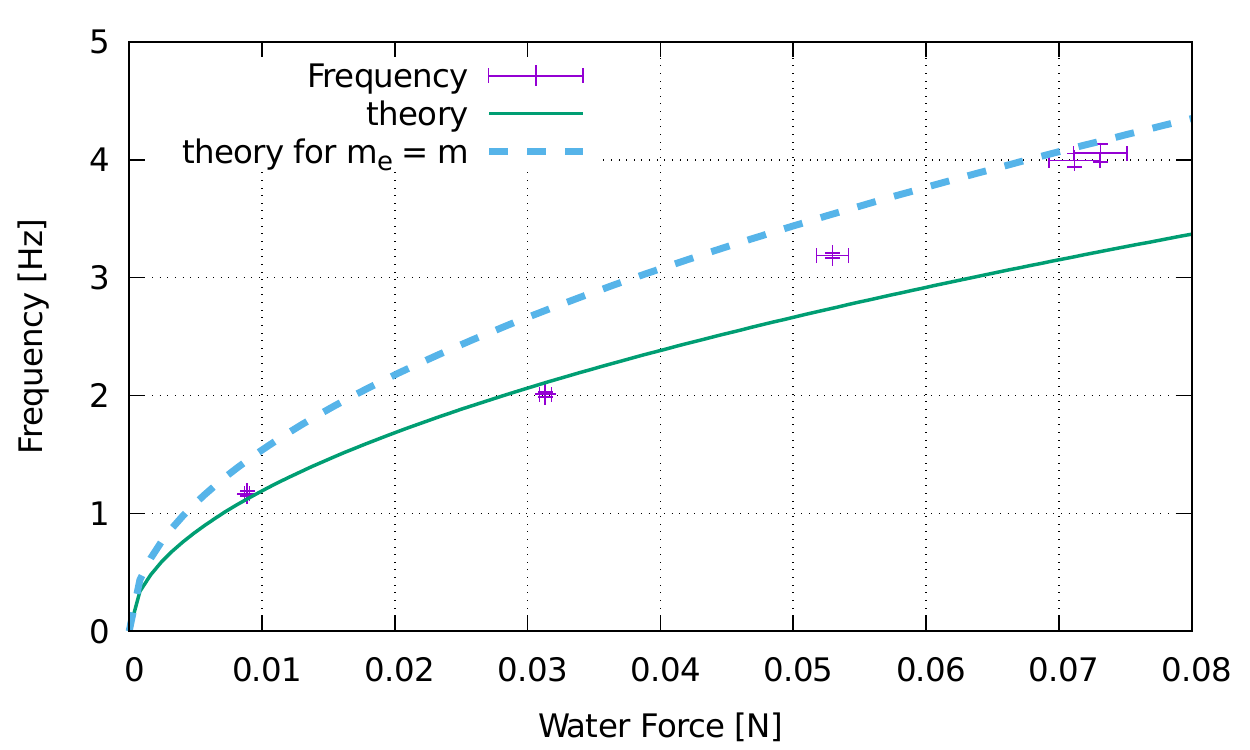}
\end{center}
\caption{The asymptotic frequency $f_{\infty}$ that is approached for long times depending on water force $F_w$. The measurement results are depicted in magenta. The theoretical predictions without slip are depicted in green. The blue dashed line represents the theory for a slipping ball using the ball mass as the effective mass.\label{fig:freq_dep}}
\end{figure}

The $k$ value from equation \ref{eq:k_definition} can be used to determine the frequency from our theory:
\begin{equation}
\frac{k}{m_e}=\frac{F_w}{m_eR} = \omega^2 \; \Rightarrow \; f_{\infty} = \frac{1}{2\,\pi} \sqrt{\frac{F_w}{m_eR}},
\end{equation}
where $m_e$ effectively denotes all inertia effects, including inertia contributions due to rotation. For our ping pong ball we assume that the mass is in a thin layer at Radius $R$. The inertia constant $m_e$ therefore is $m_e = m + \frac{\mathrm I}{R^2} = \frac{5}{3}m$. Five different water velocities have been used. The theory and the experiment are compared in Fig. \ref{fig:freq_dep}. Note there are no free parameters used in this comparison.

For small water forces $F_w$ the experimentally determined frequency (depicted violett in Fig. \ref{fig:freq_dep}) agrees with the theoretical prediction (green). However, for large forces slipping occurs, such that $m_e = \frac{5}{3}m$ no longer is a suitable assumption for the inertial contribution. For a slipping ball the inertia is mainly due to the mass, because the rotation does not occur, i.e., $m_e = m$. Therefore, in Fig. \ref{fig:freq_dep} the prediction for $m_e = m$ is shown by a blue dashed line that is in agreement with the measurements at large water forces.

\section{Conclusions and Outlook}
\label{sec:conclusions}

We studied the oscillations that occur if a ball on a flat surface is hit from above by a water jet. Obviously it is too complicated to try to describe all details of the occurring forces: Such a detailed description would require a hydrodynamic exploration of the water film that surrounds the sphere as well as various considerations concerning frictional and viscous forces and of course is far beyond what students (and even most experts in the field) can calculate. The major step towards solving the task is to find out what matters for the oscillations that should be studied. In this article we have shown that the restoring force can be understood as momentum transfer on the ball due to deflection of the water jet on the ball. However, an instantaneous restoring force could not explain any steady state with oscillations. By taking the time delay of the restoring force into account, we have shown that the time delay causes an effective driving force  leading to the observed self-sustained oscillations. Furthermore, we have compared our theoretical predictions with results from experiments.

The way how this problem has been handled is typical for problems of the IPT. An important step is to find the correct level of abstraction such that important ingredients (e.g. the time delay) are preserved while complicated details (e.g. of the hydrodynamic forces) are not considered. Students that learn to attack the problems of the IPT in such a way therefore are taught a lot about how they can handle complex research questions that they probably will deal with at later stages of their career. 

In future, the GPT will be supported by a network with locations at various German universities (see \cite{fauGPT} for details). The network was founded in december 2019 and is motivated by the successful network of the \textit{German Young Physicists' Tournament} (GYPT) which is a similar competition for high school students \cite{GYPT} and that is used as national qualification tournament for the \textit{International Young Physicists' Tournament} (IYPT) \cite{IYPT}. The goal is to learn from the success story of the GYPT and the IYPT in order to increase the visibility of the GPT and the IPT such that more students have the possibility to participate in these competitions.

\ack 
We gratefully acknowledge helpful discussions with the other members of the team from the FAU Erlangen-N\"urnberg that participated in the GPT and IPT 2019, called ``FAUltiere'': Aakash Bhat, Paul Fadler, Andreas Gramann, Lucia Härer and Janna Vischer.

\end{document}